\documentclass[%
 reprint,
superscriptaddress,
frontmatterverbose, 
 amsmath,amssymb,
 aps,
 longbibliography,
]{revtex4-1}
\usepackage{xcolor}
\usepackage{graphicx}
\usepackage{dcolumn}
\usepackage{bm}

\begin{document}

\preprint{APS/123-QED}

\title{Compact Optical Atomic Clock Based on a Two-Photon Transition in Rubidium}
\thanks{A contribution of AFRL, an agency of the US government, and not subject to copyright in the United States.}%

\author{Kyle W. Martin}
 \affiliation{Applied Technology Associates dba ATA,
 1300 Britt Street SE Albuquerque, NM 87123}
  
\author{Gretchen Phelps}%
\affiliation{%
 Air Force Research Laboratory, 
 Space Vehicles Directorate 
 Kirtland Air Force Base, NM 87117
}%

\author{Nathan D. Lemke}
\affiliation{%
 Air Force Research Laboratory, 
 Space Vehicles Directorate 
 Kirtland Air Force Base, NM 87117
}%

\author{Matthew S. Bigelow}
 \affiliation{Applied Technology Associates dba ATA,
 1300 Britt Street SE Albuquerque, NM 87123}%
 
  \author{Benjamin Stuhl}
\affiliation{%
 Space Dynamics Laboratory,
 1695 North Research Park Way  
 North Logan, UT 84341
}%
 \author{Michael Wojcik}
\affiliation{%
 Space Dynamics Laboratory,
 1695 North Research Park Way  
 North Logan, UT 84341
}%
\author{Michael Holt}
\affiliation{%
 Space Dynamics Laboratory, 
 1695 North Research Park Way  
 North Logan, UT 84341
}%

\author{Ian Coddington} 
\affiliation{%
National Institute of Standards and Technology, 
325 Broadway  
Boulder, CO 80305
}%
 
 \author{Michael W. Bishop}
\affiliation{%
 Air Force Research Laboratory,
 Space Vehicles Directorate
 Kirtland Air Force Base, NM 87117
}%

\author{John H. Burke}
\affiliation{%
 Air Force Research Laboratory,
 Space Vehicles Directorate
 Kirtland Air Force Base, NM 87117
}%

\date{\today}

\begin{abstract}
Extra-laboratory atomic clocks are necessary for a wide array of applications (e.g. satellite-based navigation and communication).  Building upon existing vapor cell and laser technologies,  we describe an optical atomic clock, designed around a simple and manufacturable architecture, that utilizes the 778~nm two-photon transition in rubidium and yields fractional frequency instabilities of $3\times10^{-13}/\sqrt{\tau (s)}$ for $\tau$ from 1~s to 10000~s.  We present a complete stability budget for this system and explore the required conditions under which  a fractional frequency instability of $1\times 10^{-15}$ can be maintained on long timescales. We provide precise characterization of the leading sensitivities to external processes including magnetic fields and fluctuations of the vapor cell temperature and 778~nm laser power. The system is constructed primarily from commercially-available components, an attractive feature from the standpoint of commercialization and deployment of optical frequency standards. 
\end{abstract}

\maketitle

\section{Introduction}

High stability clocks and oscillators play an integral role in many modern technologies such as navigation and communications \cite{Maleki2005}. Laboratory-based primary frequency standards, which utilize microwave transitions between atomic hyperfine levels, provide the highest degree of timing accuracy and are used to form international timescales \cite{Wynands2005,Parker2005}; in many cases, however, applications beyond timekeeping  require clocks that are deployed outside the laboratory setting. One well-known case is that of global navigation satellite systems (GNSS), which employ space-qualified frequency standards aboard  satellites in medium earth orbit and/or geosynchronous orbit  \cite{Maleki2005,McNeff2002}. While portable clocks are typically outpaced by their laboratory counterparts in terms of stability and accuracy, they nonetheless offer very low levels of frequency instabilities; in the case of rubidium atomic frequency standards, clocks are commercially available with a drift rate below $10^{-13}$/day and a frequency noise floor less than $10^{-14}$ \cite{Formichella2017}.

With the advent of fully stabilized optical frequency combs  in 2000 \cite{Jones2000,Diddams2000,Holzwarth2000}, optical frequency standards have rapidly  surpassed the capabilities of microwave clocks in both stability \cite{Hinkley2013,Bloom2014,Bize2016,Gill2016} and systematic uncertainty \cite{Nicholson2015,Huntemann2016,Chou2010a}. Efforts to reduce the size and increase portability of these systems is an ongoing area of interest \cite{Lisdat2017}.  However, these improvements have yet to make an impact on more stringent definitions of portable and deployable clocks. Much of the difficulty in developing compact and environmentally robust optical frequency standards lies with the complicated laser sources and optical systems required for laser cooling and interrogating an atomic sample. Moreover, given the high quality factor (i.e. narrow spectral linewidth) of typical optical clock transitions, laser pre-stabilization to a high-finesse Fabry-P\'{e}rot cavity is generally required, which adds significant complexity to the system. Finally, optical frequency combs have historically not been sufficiently compact or robust to warrant an effort toward deployment.

Fully realized frequency standards that are compact and portable, with some applications constraining total volume to less then 30 liters, while maintaining low fractional frequency instabilities, potential for less than $1\times10^{-13}/\sqrt{\tau}$, are important for a plethora of non-laboratory environments.  While microwave fountain clocks that incorporate lasers for cooling transitions \cite{Bize2012, Salomon1999} are an ongoing research effort yielding instabilities as low as  $1.4\times10^{-14}/\sqrt{\tau}$ they require the use of a cryogenic sapphire oscillator (CSO), restricting their use in non-traditional environments.  Clock compactness with low frequency instability can be maintained only through the use of an optical oscillator.   Recently, a surge of deployable microwave clocks leveraging laser cooled Rb have been integrated in satellite systems \cite{Wang2017} and others utilizing vapor cell technology and laser oscillators \cite{Emeric2017,Micalizio2015,Levi2012} have shown fractional frequency instability as low as $1.4\times10^{-13}/\sqrt{\tau}$, with potential to meet constrained size and power requirements for on orbit operation.

\begin{figure}
\includegraphics[width=0.47\textwidth]{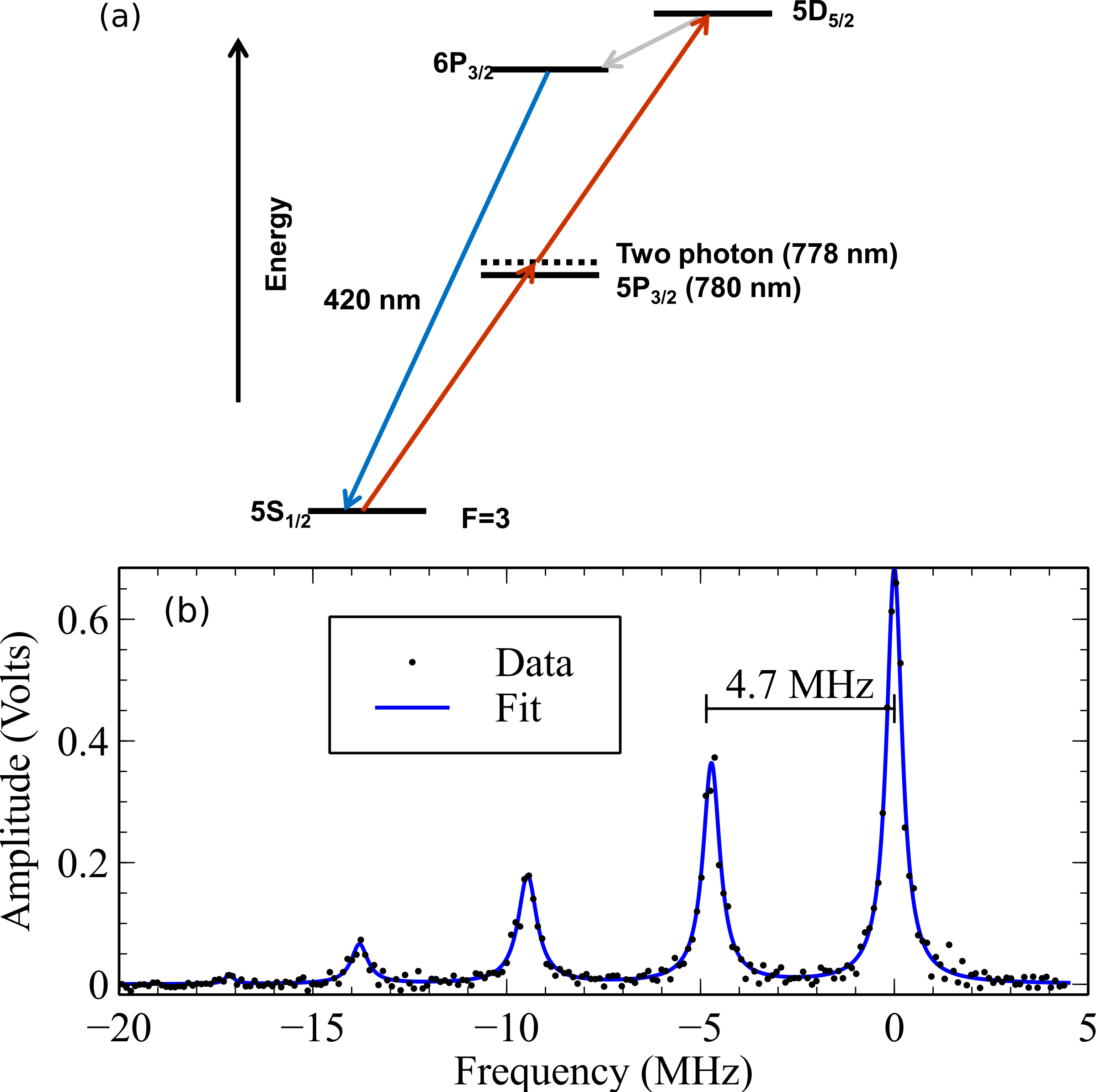}
\caption{\label{fig:twophoton} (a) Partial energy level diagram of Rb. The virtual state associated with two-photon excitation at 778~nm is shown by the dashed line.  The cascade decay path 5D$_{5/2}\rightarrow$ 6P$_{3/2} \rightarrow$ 5S$_{1/2}$ results in the emission of a 420~nm photon, which we use to observe the two-photon resonance. (b)~Two-photon excitation spectrum of $^{87}$Rb (F=2). The frequency axis is presented as the 778~nm laser detuning from the F=2$\rightarrow$F'=4 transition, which has the largest Clebsch-Gordan coefficient. Included is a fit using a sum of four Lorentzian peaks of equal width and with relative peak heights and detunings constrained by the values presented in \cite{Nez1993}. The fitting procedure results in a full-width at half-maximum linewidth of 609 kHz, which exceeds the natural linewidth by a factor of 1.8, which we expect is due to line broadening from helium collisions \cite{Zameroski2014}.} 
\end{figure}

With these challenges in mind, we investigate an optical clock based on a two-photon transition in a hot Rb vapor. Two-photon transitions are attractive because they enable Doppler-free spectra \cite{Grynberg1977} without the need for laser-cooling, provided two anti-parallel laser beams are used to interrogate the atomic vapor. Moreover, the two-photon transition can often be observed via a fluorescence signal that is spectrally resolvable from the probe laser; together with the large number of atoms interrogated in the vapor phase, this enables a very high signal-to-noise measurement of the clock transition.  For the case of the Rb two-photon transition at 778~nm, fluorescence is readily observable at 420~nm, and stray light in the near infrared is rejected with standard optical filtering. We further benefit in the case of the Rb~5S$_{1/2}\rightarrow$~5D$_{5/2}$ transition from the nearby intermediate state 5P$_{3/2}$ that is only separated by 2~nm from the virtual two-photon state (see Figure~\ref{fig:twophoton}), facilitating significant atomic excitation rates at modest optical intensities \cite{Nez1993,Touahri1997,Hilico1998}. Conveniently, 778.1~nm light can be produced by second harmonic generation (SHG) of 1556.2~nm, which falls in the telecommunications C-band, allowing the use of mature laser sources and erbium fiber frequency combs \cite{Kuse2015,Lezius2016,Sinclair2015}. Of particular benefit is the availability of commercial laser systems for which the fast linewidth is significantly below the natural linewidth of the excited clock state ($\Delta \nu \approx 330$~kHz as observed at 778~nm), alleviating the requirement for laser pre-stabilization to a high finesse optical cavity.  There are, however, remaining challenges that persist in hot atomic vapor, two-photon clocks.  Two-photon transitions have large AC Stark shifts which must be mitigated or compensated and precision temperature control is required while probing the hot atomic vapor.  These two major problems need to be addressed for any suitable frequency standard based on this transition.  

The appealing features motivate our investigation of an optical Rb atomic frequency standard (``O-RAFS'') as a future portable clock. While our work has not yet specifically designed the system for a small footprint and low power consumption, we anticipate this should be possible. As a first step, however, it must be shown that a vapor cell-based two-photon clock can meet the demanding timing requirements needed to enable the above applications. Of particular interest is a frequency standard that can surpass existing microwave clocks by one factor of 10 in both short- and long-term stabilities , which would translate to an Allan deviation of $\sim 1 \times 10^{-13}$ at 1~s and  $\sim 1 \times 10^{-15}$ at 1~day \cite{Schuldt2017}. Previous investigations into  this transition in Rb were carried out in the 1990s \cite{Nez1993,Touahri1997,Hilico1998} with some renewed interest recently leveraging the development of laser and frequency comb technologies \cite{Marian2004,Wu2014,Rathod2015}. While initial experiments showed the system to be capable of instabilities as low as $3\times 10^{-13}/\sqrt{\tau}$ for $\tau$ from 1--2000 s, on longer timescales the instability increased \cite{Hilico1998}, and instability below $1\times 10^{-14}$ has not been observed. Here, we  extend the range of integration to longer timescales and demonstrate a corresponding reduction of long-term instability, approaching the level of $1\times 10^{-15}$. Key to this level of performance are tight control over the vapor density \cite{Zameroski2014} and laser power \cite{Hilico1998}, both of which we describe in detail below, together with a full stability budget for the frequency standard.

This paper is organized as follows.
 
Section \ref{Sec:Experiment} describes the design of the frequency standard and measurement system. Section~\ref{Sec:Instabilities} provides detailed analysis of the known contributors to clock instability, with particular emphasis on the clock laser-induced AC Stark shift, collisional effects, Zeeman shifts, and short term stability limitations. The performance of the frequency standard is detailed in Section~\ref{Sec:Results}, and  the paper concludes with a discussion of conceived upgrades to the two-photon frequency standard in Section~\ref{Sec:Conclusion}.

\section{Experimental apparatus}\label{Sec:Experiment}

\begin{figure*}
\includegraphics[width=\textwidth]{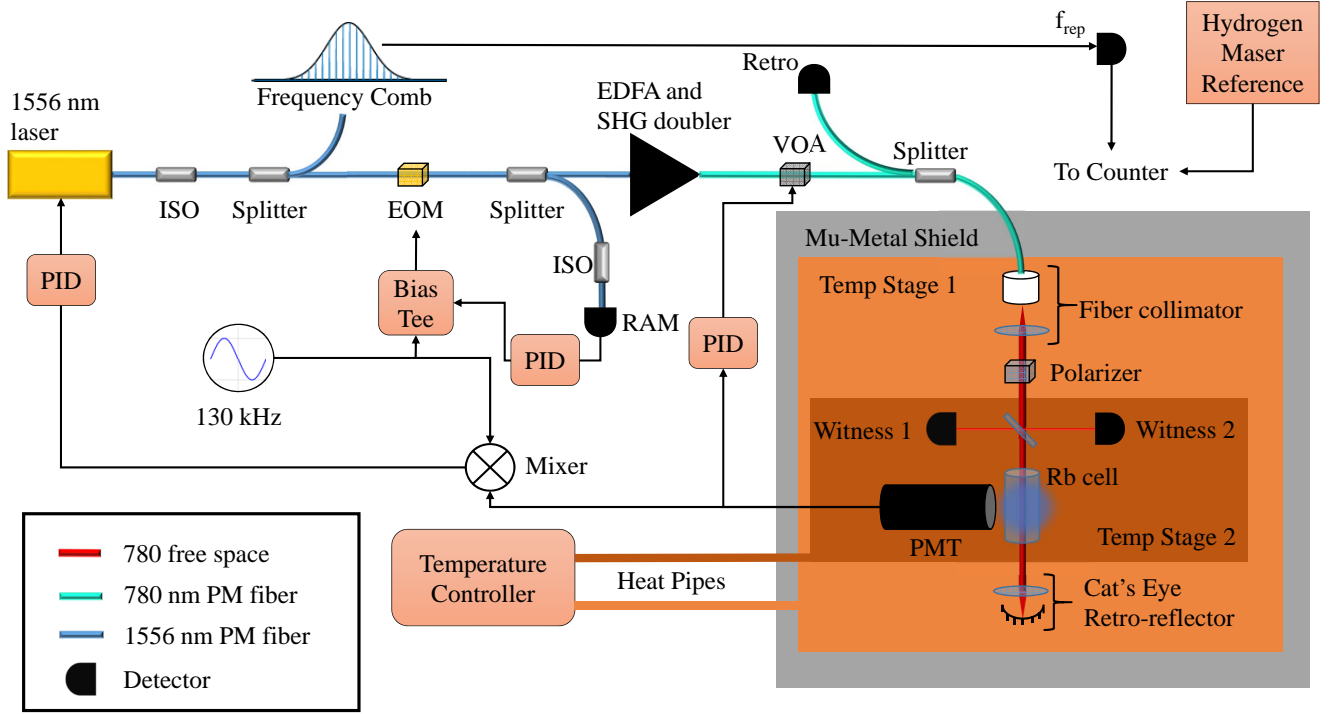}
\caption{\label{fig:quantel} An optical and simplified electrical schematic of the Rb two-photon frequency standard as described in the text. EOM - electro-optic modulator; PMT - photomultiplier tube;  VOA - variable optical attenuator; ISO - optical isolator; EDFA - erbium doped fiber amplifier; SHG - second harmonic generator; RAM - residual amplitude modulation; PID - proportional integral differential lock mechanism.}
\end{figure*}

The design shown in Figure \ref{fig:quantel} begins with a 1556~nm diode laser system  that produces 
20~mW of narrow-band light, with a fast linewidth $\leq 2$~kHz. This laser acts as the local oscillator for the frequency standard. A small portion of the seed laser output is sampled by a fiber splitter to form an optical beat note with a fiber frequency comb based upon the design in \cite{Sinclair2015}. The fully self-referenced frequency comb coherently divides the  385~THz optical waveform to $\sim$~200~MHz, which is the comb's pulse repetition rate.  After stabilization of the optical beatnote and carrier envelope offset frequency, the comb's repetition rate is photodetected and a Microsemi 5125A is used as a frequency counter and the phase noise is compared to a hydrogen maser. The remaining portion of the 1556~nm laser output enters a fiber-coupled electro-optic modulator (EOM) formed in a proton-exchange waveguide embedded in lithium niobate, which is driven at 130~kHz. After the EOM, the light is amplified by an erbium-doped fiber amplifier and undergoes second harmonic generation (SHG) in a PPLN crystal, outputting as much as 1 W of 778.1~nm light. The output of the SHG crystal, typically around 100 mW, is subsequently sent through a variable optical attenuator (VOA), which is used for laser power stabilization as described below.  Typically 30~mW of 778~nm light is delivered to the vapor cell assembly. 


The vapor cell assembly  is enclosed in a 5~mm-thick, single layer mu-metal magnetic shield, to reduce spectral broadening associated with the Zeeman shift. The vapor cell is heated to 100~$^{\circ}$C to generate sufficient vapor density for a high stability clock. To avoid local magnetic fields when heating the vapor cell, all heat is generated with resistive and thermo-electric devices located outside the magnetic shielding; water-filled heat pipes protrude through the magnetic shield and provide heat to the dual-zone temperature control stage surrounding the vapor cell. The vapor cell, which is a rectangular parallelepiped with dimensions 5$\times$5$\times$25~mm, containing $>99\%$ isotopically enriched $^{87}$Rb, is placed such that it has a $1~$K thermal gradient along its length, forcing the cell's cold spot on the pinched-off fill tube of the borosilicate glass cell.  The vapor cell is oriented at Brewster's angle with respect to the incident laser beam to reduce stray reflections. 

The 778~nm laser output is delivered by polarization-maintaining optical fiber through an opening in the magnetic shield, where it is collimated (1/e$^2$ intensity radius $w_0 = 0.66$~mm) using a non-magnetic optical assembly.  A calcite Glan-Taylor polarizer is placed at the output of the fiber launcher to reduce polarization wander.  The laser beam is sampled by a glass plate pick-off before entering the vapor cell; un-modified Thorlabs SM05PD1A photodetectors, with dark current 20~nA and square active area of 3.5 mm on a side, on each side of the glass plate monitor the optical power in the sampled beams, which remain relatively collimated with waist of 0.66~mm.  A cat's eye retro-reflector \cite{Snyder1975} provides a precisely  anti-parallel reflected beam, which is necessary for eliminating  Doppler broadening.  A portion of the fluorescence at 420~nm passes through a short-pass optical filter and is detected by a photomultiplier tube (PMT), dark current 5~nA. The 
PMT provides a fast temporal response and high electron-multiplying gain. After a transimpedance amplifier, the PMT output signal is demodulated by the 130~kHz sinusoidal modulation applied to the EOM in a phase detector, resulting in a laser detuning-dependent error signal. A digital servo controller with dual integrators and approximately 50~kHz bandwidth, feeds the 1556~nm laser's current to hold the laser on the two-photon resonance.

This design allows for the study of various parameters that contribute to the system's performance at different time scales.  The short-term stability is determined by the atomic linewidth, optical intensity, detector collection efficiency, and laser frequency noise characteristics. The long-term stability, with a current goal of $ < 1 \times 10^{-15}$ at one day, requires the stabilization of various experimental and environmental parameters including the vapor cell temperature (Rb vapor density), magnetic field, and optical power; these parameters are investigated in subsequent sections.

\section{Sources of clock instability}\label{Sec:Instabilities}

In this section we discuss the leading sources of instability to the Rb two-photon system.  Particular importance is paid to rigorously determining the relevant sensitivity coefficients. Because our goals call for a stable frequency standard but not necessarily one with high accuracy, we do not undertake to precisely measure the magnitude of each systematic effect, but rather to characterize the stability requirements of external parameters such as magnetic field and laser power. 
Table \ref{tab:table2} summarizes all of the clock shifts and  environmental stability parameters necessary to achieve fractional frequency instabilities of  $1 \times 10^{-15}$. 

\subsection{AC Stark shift}

\begin{figure} 
\includegraphics[width=0.47\textwidth]{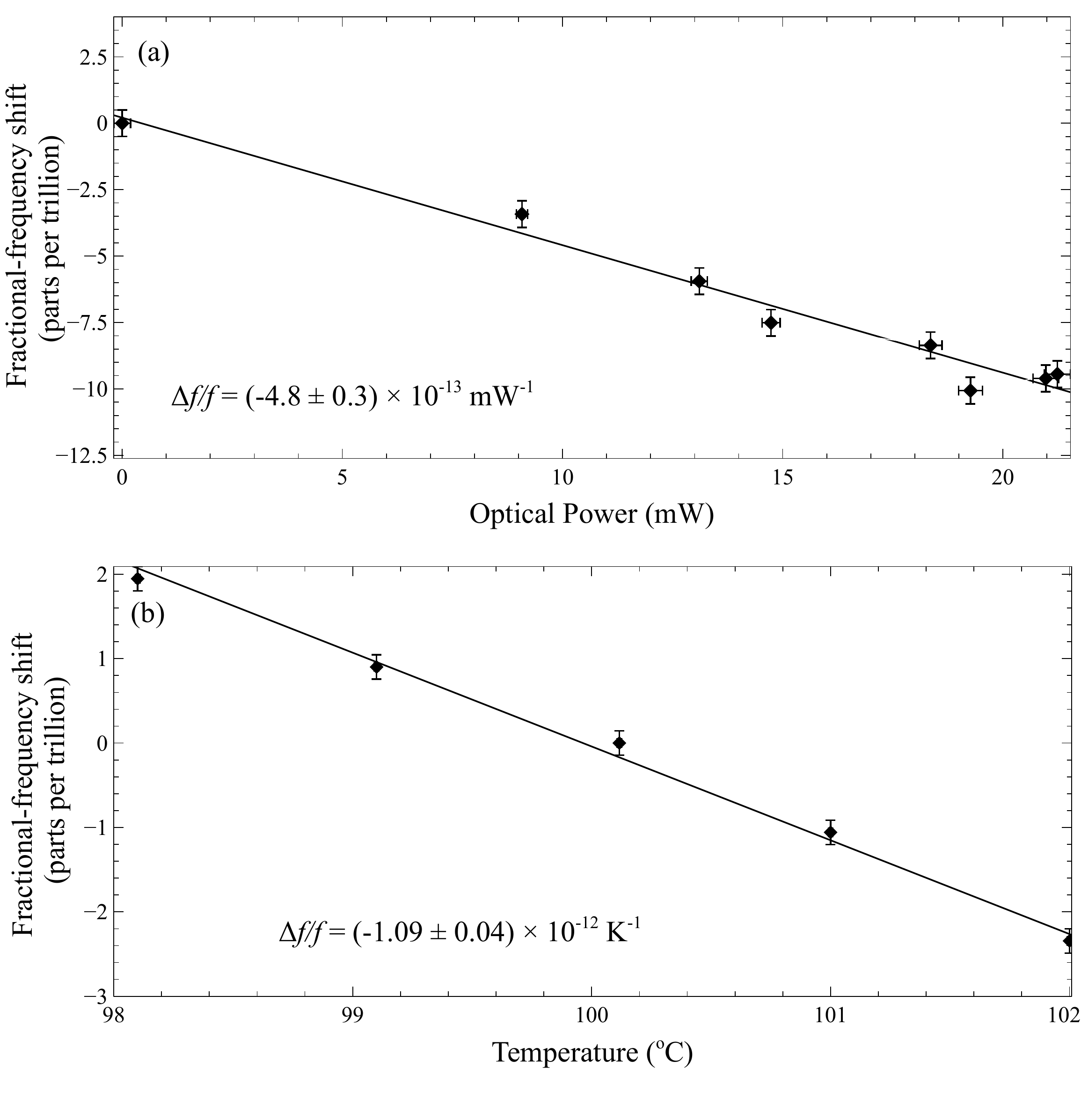}
\caption{\label{fig:light_shift} (a) Experimentally measured 778~nm AC Stark shift for a (0.66 $\pm$ 0.05) mm beam.  The fit used was orthogonal distance regression (ODR) which weights error bars in x and y data yielding a reduced $\chi^2$ of 1.57 (b) Experimentally measured Rb collisional shifts, fit with ODR yielding a reduced $\chi^2$ of 0.908.}
\end{figure}

Two-photon transitions are well-known to suffer from sizable AC Stark shifts associated with the probe laser \cite{Grynberg1977}. The fractional AC Stark  shift is given by 

\begin{equation} \label{equation:stark1}
\frac{\Delta \nu}{\nu_0} = \frac{\Delta \alpha}{2c\epsilon_0 h} \overline{I \left(r\right)} = k\left(w_0\right) P,
\end{equation}
where $\overline{I \left (r \right )} \propto P/w_0^2$ is the spatially-averaged laser intensity, $P$ is the one-way optical power incident on the vapor, $\nu_0 \approx$~385~THz is the two-photon laser frequency, $w_0$ is the $1/e^2$ intensity radius, $\Delta \alpha$ is the differential polarizability of the two clock states at 778.1~nm and $c$, $\epsilon_0$ and $h$ are the speed of light, permittivity of free space and Planck's constant respectively. We measured the shift experimentally utilizing the clock laser detailed in Figure \ref{fig:quantel} together with an external Ti:sapphire laser.  The Ti:sapphire laser was tuned slightly away  from the two-photon resonance by 2.6~GHz to an optical frequency of  385287.8~GHz, far enough detuned to induce no measurable excitation of the vapor, yet near enough to not significantly change the polarizability. The two lasers were combined by a 50:50 beamsplitter and coupled into a single mode fiber, thereby enforcing the same spatial mode. Without changing the florescence signal size, which would contaminate the Stark shift measurement via lock point fluctuations, we varied the power of the detuned laser and measured the associated shift.

The results of this measurement are shown in Figure~\ref{fig:light_shift} along with a linear regression used to determine the sensitivity coefficient $k\left(w_0\right)$.  The measured fractional clock shift coefficient is 4.8(4)$\times 10^{-13}$/mW for $w_0=0.66(5) \mu$m. 

A previous measurement \cite{Hilico1998}, appropriately scaled to match our  beam radius, reports a coefficient of 4.5(4)$\times 10^{-13}$/mW, which agrees well within the error bars of the two measurements.

This coefficient indicates that the optical power must be stabilized to 2.1~$\mu$W  to achieve $1 \times 10^{-15}$ clock instability, requiring a precise laser power controller.  A  laser power stabilization circuit was constructed  (Figure~\ref{fig:quantel}) using feedback to a fiber-optic variable optical attenuator, which supports a loop bandwidth of 1~kHz.  We found it most effective to use the fluorescence signal detected on the photomultiplier tube as the laser power sensor, rather than a sampled beam measured on a photodiode, although the latter is used as an out-of-loop witness sensor. This out-of-loop data was used to determine the fractional clock limitation imposed by laser power instability as shown in Figure~\ref{fig:adev2}.

\subsection{Zeeman shift}

\begin{figure} 
\includegraphics[width=0.47\textwidth]{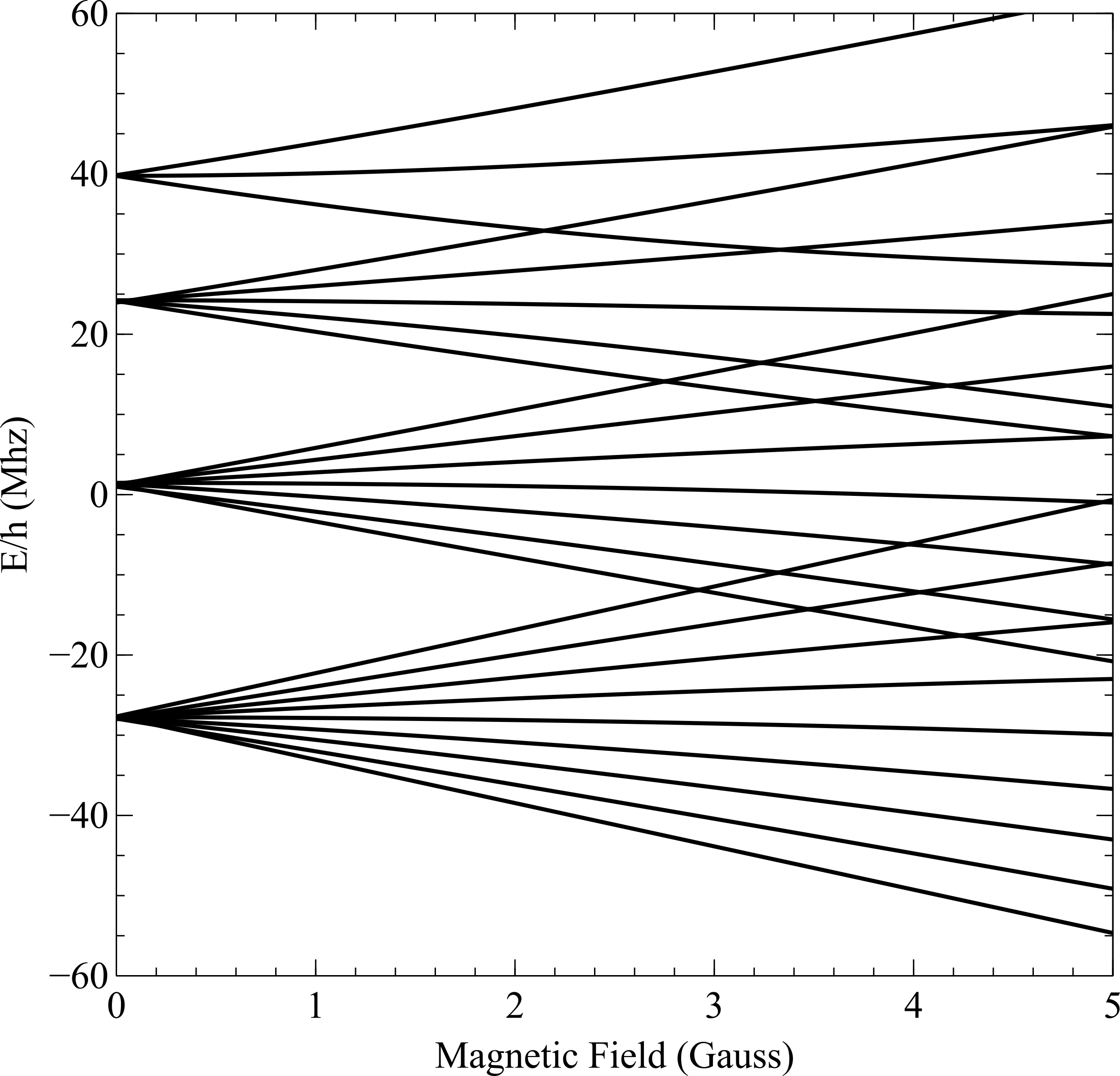}
\caption{\label{fig:B-field} Magnetic  field splitting for the $5D_{5/2}$ states of of $^{87}$Rb, which was determined by numerical diagonalization of the total Hamiltonian as described in the text.}
\end{figure}

Stray magnetic fields are an important environmental variable that can produce substantial atomic frequency shifts.   The magnetic field shift in the incomplete Paschen-Bach regime of the $5S_{1/2}$ ground state can be analytically calculated utilizing the Briet-Rabi formula; because our spectroscopic technique does not resolve transitions between specific magnetic sublevels, we then average over all relevant $m_I$ and $m_J$ magnetic quantum numbers. This assumption results in no first-order (linear) dependence of the clock frequency on magnetic field, and is valid for local magnetic fields $\lesssim$100~mG, which is roughly the field at which Zeeman-induced line-broadening  exceeds the natural linewidth of the two-photon transition. 
Substitution of the Land\'{e} g-factors, $g_J$ and $g_I$  and the magnetic dipole constant from Refs. \cite{Arimondo1977,Salomon1999} yields a second order state shift of $114$~Hz/G$^2$ for $^{87}$Rb (F=2) and $358$~ Hz/G$^2$ for $^{85}$Rb (F=3). 
The clock shift for the $^5D_{5/2}$ excited state does not have a simple analytical solution.  The Hamiltonian,
\begin{align*}
&\boldsymbol{H}=\boldsymbol{H}_{hfs}+\boldsymbol{H}_{B}^{(hfs)},\\
&\boldsymbol{H}_{hfs}=A_{hfs}\frac{\boldsymbol{I}\cdot \boldsymbol{J}}{\hbar^2}\\
&+B_{hfs}\frac{\frac{3}{\hbar^2}(\boldsymbol{I}\cdot \boldsymbol{J})^2+\frac{3}{2\hbar}(\boldsymbol{I}\cdot \boldsymbol{J})-J(J+1)I(I+1)}{2I(2I-1)J(2J-1)},\\
&\boldsymbol{H}_{B}^{(hfs)}=\mu_B (g_J \boldsymbol{J_z}+g_I \boldsymbol{I_Z})B,
\end{align*}
where, $\boldsymbol{I}$ is the nuclear spin, $\boldsymbol{J}$ is the total angular momentum, $\mu_B$ is the Bohr magneton and B is the applied magnetic field,  was generated and  diagonalized numerically. Substituting the magnetic dipole and quadrapole constants, $A_{hfs}$ and $B_{hfs}$, for the $5D_{5/2}$ level from Ref. \cite{Nez1993} results in a state shift of  $50$~kHz/G$^2$ for $^{87}$Rb (F=4) and $190$~kHz/G$^2$ for $^{85}$Rb (F=5). Expressing the differential Zeeman sensitivities in fractional frequency units, we find the net clock shifts to be $6.5 \times 10^{-11}$/G$^2$ for $^{87}$Rb and $2.5 \times 10^{-10}$/G$^2$ for $^{85}$Rb.  The  energy level splitting diagram for the exited state of $^{87}$Rb is shown in Figure \ref{fig:B-field}.  With these coefficients, we can now specify the magnetic shielding requirements; for $^{87}Rb$ ($^{85}Rb$), the magnetic field should be stable at the 3.9~mG (2.0~mG) level. We installed a rectangular $\mu$-metal shield of thickness 5~mm for which the expected shielding factor exceeds 1000. In practice, the shielding factor is reduced to due to openings for the heat pipes, optical fiber, electrical cabling, and photomultiplier tube, but we nonetheless expect the residual magnetic field at the vapor cell to be $\lesssim$ 1~mG.

\subsection{Collisional shift}

The vapor cell temperature was determined using a standard 100 ohm resistive temperature detector (RTD) four wire measurement, with a duplicate device for out-of-loop monitoring.  Two independent temperature control stages were designed, a 333~K plate (temperature stage~1 in Figure~\ref{fig:quantel}), to provide a stable reference temperature for heat transfer control,  and a second, more finely controlled 373~K stage (temperature stage~2) upon which the vapor cell was mounted (see Figure \ref{fig:quantel}).   These stages were separated by four G-11 fiberglass posts to provide conductive thermal isolation. A precision temperature controller regulated a thermoelectric device and closed the temperature servo loops. Fiberglass insulation was added around the temperature control stages to reduce convective heat loss.

The collisional shift of Rydberg states of alkali metals has been a subject of research since E. Fermi formulated a pseudo-potential to describe the interaction of the atoms loosely bound electron with the perturber \cite{Fermi2008}. There has been a large disagreement in the literature about the measurement of the collisional shift \cite{Osama2016,Zameroski2014,Hilico1998}.  We measured a collisional shift for our $^{87}$Rb enriched vapor cell by varying the vapor cell temperature and measuring the clock shift of our locked laser system.  For each temperature change the system was allowed to thermalize, whereupon the resulting frequency shift was measured over 300 s.

The obtained frequency vs. temperature is plotted in Figure \ref{fig:light_shift}, and the fit supports a clock shift of $-1.09(4)\times10^{-12}$/K, which is a factor of 2 larger than reported for $^{85}$Rb by Zameroski \textit{et al.} \cite{Zameroski2014} in a vapor cell with natural Rb.  
At 373~K the temperature must be stable to $0.92$ mK to achieve fractional frequency stability of $1 \times 10^{-15}$.  
The fractional clock limitation caused by the temperature fluctuations of temperature stage~2, measured by the out-of-loop RTD, is shown in Figure \ref{fig:adev2}.

\subsection{Other Considerations}

A frequency modulation technique utilizing a phase modulator is employed to lock the laser to the Rb~5S$_{1/2}\rightarrow$~5D$_{5/2}$ two photon transition.  This technique is known to suffer from residual amplitude modulation  (RAM) that arises when modulation sidebands are not equal in magnitude or opposite in phase \cite{Bjorklund1980, Jun1998}.  Zhang \textit{et al.} developed a technique to suppress both in-phase and quadrature RAM  \cite{Zhang2014} utilizing a feedback control of the phase modulator's DC bias and temperature, respectively. 
We employed a similar technique using a single feedback loop to the DC bias voltage, supporting a loop bandwidth of 10~kHz, which was combined with a sinusoidal modulation signal on a bias tee as shown in Figure~\ref{fig:quantel}. This method yielded   suppression of $> 35$~dB. Additionally, we find benefit in saturating the input to the 1556~nm optical amplifier, which provides a passive reduction of RAM of $> 5$~dB. While these two suppression mechanisms were sufficient to achieve fractional clock instabilities shown in Figure~\ref{fig:adev2}, further corrections to quadrature RAM could be implemented through stabilizing the temperature of the EOM to further decrease clock instabilities.

Doppler effects are largely eliminated by retroreflecting the laser beam that passes through the vapor cell. However,   residual broadening related to the  absorption of two-photons from the same beam remains; this contribution to the lineshape is a Gaussian function with a full-width at half-maximum of $\sqrt{8 k_B T \ln{2}/mc^2}\nu_0 \approx$ 571~MHz for $^{87}$Rb at $T=$373~K, with $k_B$ the Boltzmann constant, and $m$ the atomic mass.  Absorbing two photons from the same beam occurs with the same probability as absorbing one photon from each beam; however, the linewidth associated with the former process is 1000 times greater than the latter.  Hence, the  Doppler-broadened peak is not easily resolved, and residual Doppler effects are small.

The significant tails of the Lorentzian peaks of neighboring hyperfine transitions pull the spectral lines closer together, a phenomenon known  as line-pulling.  The amount by which a particular transition is shifted is calculated by summing over all relevant hyperfine Lorentzians with appropriate frequencies and strengths given by Ref.~\cite{Nez1993}.  The two-photon transition is shifted by 0.477~Hz for $^{85}$Rb and 0.030 Hz for $^{87}$Rb.

Second-order Doppler broadening, taking into account first order relativistic corrections, is given by, 
\begin{equation}
\frac{\delta \omega}{\omega} = \frac{\bar{v}^2}{2c^2},
\end{equation}
where  $\bar{v}^2=8k_bT/m\pi$.   For Rb at 373~K the fractional clock shift is $5 \times 10^{-13}$ with a slope of $1 \times 10^{-15}$/K.

The atomic vapor is immersed in a bath of electromagnetic radiation whose spectrum follows Planck's Law.  In many cases, the blackbody radiation (BBR) shift can be treated as a DC Stark shift, since the radiation is far off resonance from all relevant atomic transitions \cite{Itano1982}.  However, the operational temperature of our system, 373 K, yields a blackbody spectrum that is nearly resonant with several transitions connecting to the $5D_{5/2}$ state.  Farley and Wing derived this perturbation for hydrogen, helium and the alkali-metal atoms for electromagnetic radiation at 300 K \cite{Farley1981}.  Hilico \textit{et al}.  calculated, assuming a $T^4$ behavior, that the perturbation would yield a shift of --210~Hz with a local slope of $\sim$1~Hz/K \cite{Hilico1998} at 373~K.  The fractional clock shift arising from BBR is $1.3 \times 10^{-15}$/K requiring that the blackbody source be held to temperatures more stable than 770~mK; however, atomic Rb transitions near resonance with the blackbody spectrum at 373 K possibly impact the accuracy of this estimate.  

The D.C. polarizability of the $5D_{5/2}$ state  was measured in \cite{Tregubov2015}, and it exceeds that of the $5S_{1/2}$ state by a factor of $\sim$ 50 due to low-lying transitions to nearby levels. Using this polarizability, we calculate the fractional clock sensitivity to D.C. electric fields is $5.9\times 10^{-15}$/(V/cm)$^2$. The magnetic shield surrounding the vapor cell assembly also acts as a Faraday cage to prevent external electric fields from reaching the atomic vapor. However, stray charge could accumulate on the glass vapor cell itself; any resulting patch potentials need to be stable at the  0.5~V level.

Experimentally determined collisional shifts in \cite{Zameroski2014} for various noble gases were examined to put limits on vapor cell impurities. Helium is the only gas known to permeate the vapor cell, and it produces frequency shifts of --2.1~MHz/Torr.  Therefore, we obtain that a helium leak rate of $< 3.6\times 10^{-8} $ Torr/day must be achieved in order to achieve fractional clock instabilities below $1 \times 10^{-15}$. The cell may also be permeable to methane, which has an atmospheric composition of about three times less than helium, but the shift rate due to methane has not been measured to our knowledge.

\begin{table}
\caption{\label{tab:table2} The environmental variables that impact $^{87}$Rb clock performance are listed along with the corresponding fractional frequency sensitivity coefficient. The right column tabulates the stability requirement for each parameter to support a fractional frequency instability of  $1\times10^{-15}$.}
\begin{ruledtabular}
\begin{tabular}{lcr}
Shift&Fractional&Stability \\ 
 &Coefficient  & at one day \\
\hline \\

778 nm AC Stark & $4.8 \times 10^{-13}$/mW & 2.1$~\mu$W\\
Rb density & $1.1\times 10^{-12}$/K & 0.92 mK\\
Blackbody Radiation & $1.3\times 10^{-15}/K$ & 770 mK\\
DC Stark & $5.9\times 10^{-15}$/(V/cm)$^2$ & 0.17 (V/cm)$^2$\\
2$^{nd}$ Order Doppler & $1.0\times 10^{-15}/K$ & 1.0 K\\
Zeeman & $6.5 \times 10^{-11}$/G$^2$ &3.9 mG \\
Helium Collisional& $2.7\times10^{-8} /$Torr   & $3.6\times 10^{-8} $ Torr \\
\end{tabular}
\end{ruledtabular}
\end{table}

\subsection{Short Term Stability}

The practical noise limit of a frequency standard is the greater of the local oscillator noise and the shot noise limit of the atoms or the photons used to detect those atoms.  
The Allan deviation, limited by shot noise can be written, 
\begin{equation}
\sigma_y^{(SN)} = \frac{1}{\nu_0} \sqrt\frac{S_f}{2\tau},
\end{equation}
where, 
\begin{equation}
S_f = \left(\frac{g}{p}\right)^2\frac{S_v}{2},
\end{equation}
g is the mixer gain, p is the error signal slope in Hz/V, $S_v$ is the voltage spectral density and $\nu_0$ is the transition frequency \cite{Hilico1998}.  For the described system the necessary parameters to calculate the shot noise limit are shown in Table \ref{tab:shotnoise} and yield a shot noise limit of $4.6\times 10^{-13}/\sqrt{\tau}$ for a 10~mW of light and $2.7\times 10^{-13}/\sqrt{\tau}$ for a 30~mW beam.
\begin{table}

\caption{\label{tab:shotnoise}Signal parameters for the ORAFS system with 10~mW of input power.}
\begin{ruledtabular} 
\begin{tabular}{lr}
parameter&value \\ 
\hline \\
Mixer gain (g)& 0.41\\
Signal to noise in 10 kHz bandwidth & 21.7~kHz \\
Error signal slope (p)& 9.56$\times10^{-8}$ V/Hz\\
Voltage spectral density ($S_v$)& 6.9$\times10^{-9}$V$^2$/Hz\\
Power spectral density at 500 kHz & 2.7$\times10^{-27}$ Hz/Hz$^2$\\
\end{tabular}
\end{ruledtabular}
\end{table}

Although the current system is not shot noise limited the fundamental limit set by the clock laser can be written as,
\begin{equation}
\sigma_y^{(SN)} = \frac{S_y^{(LO)}[2f_m]}{ 2 \sqrt{\tau}},
\end{equation}
where, $f_m$ is the modulation frequency, $S_y^{(LO)}$ is the power spectral density of the local oscillator's fractional frequency noise \cite{Audoin1990}.  The power spectral density of the seed laser used at twice the modulation frequency yields a limit of $2.6\times 10^{-14}/\sqrt{\tau}$, shown in Figure \ref{fig:adev3}. 

In principle the short term instability of the ORAFS system could be decreased by either collecting more florescence or turning up incident laser power.  Figure \ref{fig:adev3} shows an instance where fluorescence collection and laser power were simultaneously increased, SNR in a 10~kHz bandwidth increased to 210 kHz.  The predicted shot noise limit for this data set was $8\times 10^{-14}/\sqrt{\tau}$.  Unfortunately, after a short time the clock begin to drift from large AC Stark shifts.  

Efforts towards increasing florescence detection are ongoing, however, increasing laser power remains an unappealing option, as that would yield larger AC Stark shifts, already a daunting task to mitigate.

\section{Results} \label{Sec:Results}

\begin{figure} 
\includegraphics[width=0.5\textwidth]{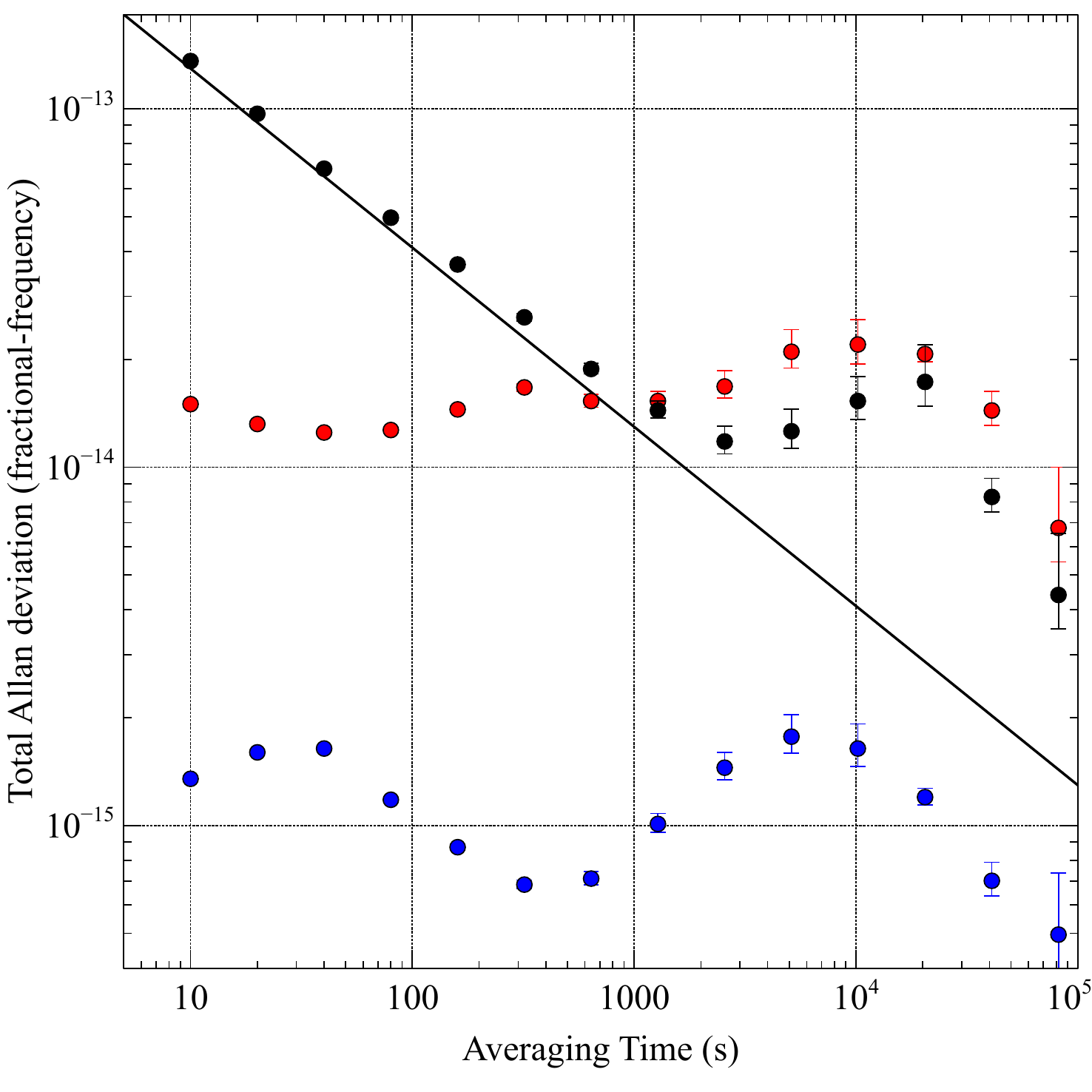}
\caption{\label{fig:adev2} The  fractional frequency instability plotted as a Total Allan deviation for $^{87}$Rb with $1/\sqrt{\tau}$ white noise (black) as well as anticipated limits on the clock stability arising from cell temperature fluctuations (blue) and laser power fluctuations (red).  We believe the instability limit arising from laser power fluctuations is an overestimate due to temperature-dependent effects in the witness photodiode as described in the main text}
\end{figure}

\begin{figure} 
\includegraphics[width=0.5\textwidth]{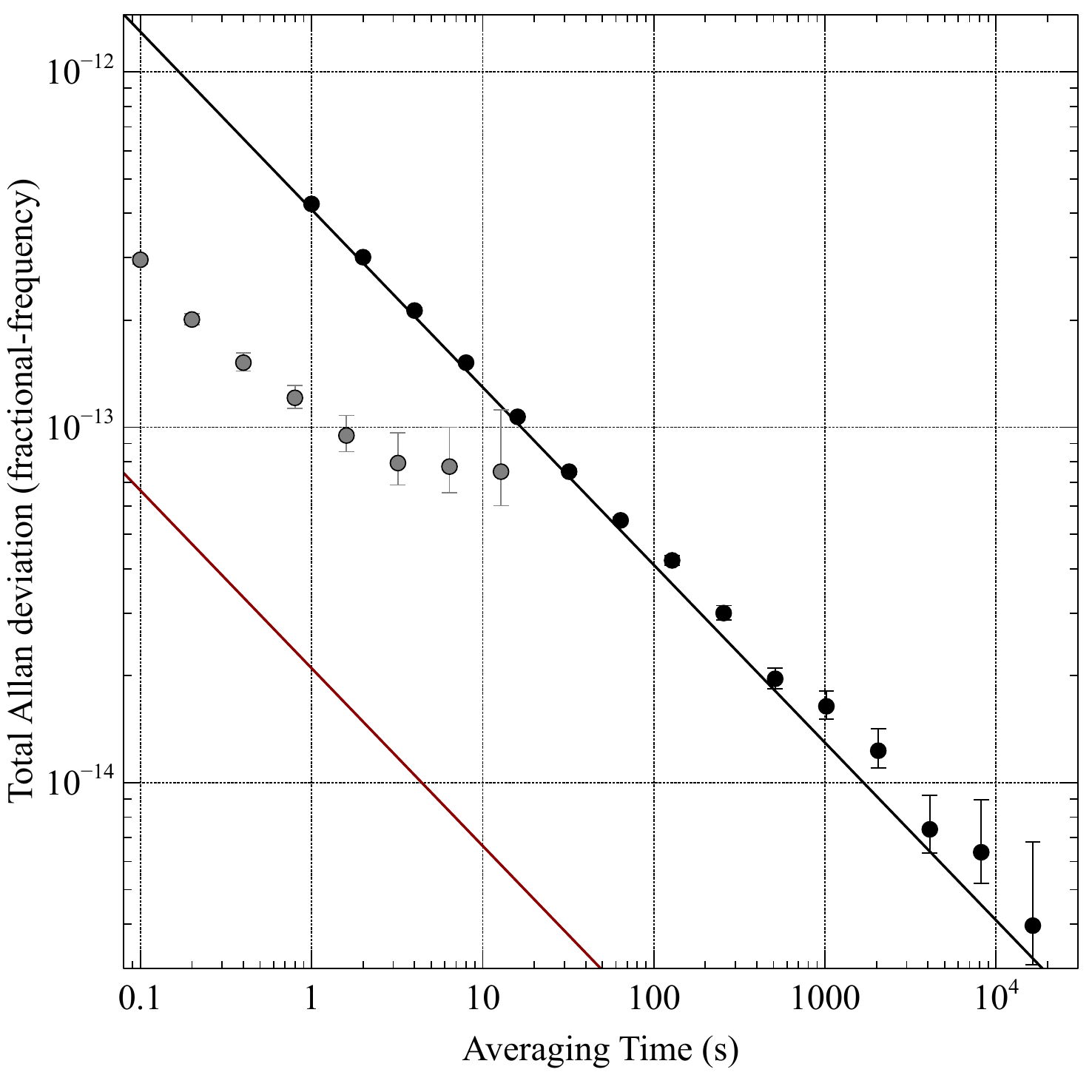}
\caption{\label{fig:adev3}  The  fractional frequency instability plotted as a Total Allan deviation for $^{87}$Rb with $1/\sqrt{\tau}$ white noise  for a smaller section of the data presented in Figure \ref{fig:adev2} (black).  Also shown: A second data set with a larger signal to noise ratio (grey) and the intermodulation limit for the laser system described in Section \ref{Sec:Experiment} (red).   }
\end{figure}

Having assessed the leading contributions to instability, we next measured the clock performance by collecting the  comb repetition rate and compared the phase noise to a hydrogen maser. During data collection, the vapor cell temperature and 778~nm laser power were monitored. The phase comparison was sampled at a rate of 1~Hz before being converted to frequency data, from which a linear drift of $-1 \times 10^{-18}$/s was removed. Figure \ref{fig:adev2} shows the resulting total modified Allan deviation of the system, as well as the expected clock performance limitations derived from out-of-loop measurements of the cell temperature and laser power. Clock performance exceeds expected stability as calculated from laser power measurements, however, long term laser power measurements from the witness photodiode are thought to be partially influenced by room temperature fluctuations; these temperature variations lead to an overestimate of Stark shift-related clock instability. Figure \ref{fig:adev3} examines a portion of the same data, highlighted because of relative instability of the light shift was low for a period.  As shown in Figure \ref{fig:adev3}, the Rb two-photon frequency standard can operate with a fractional frequency instability of $3\times10^{-13}/\sqrt{\tau (s)}$ for $\tau$ from 1~s to 10,000~s, however, complications with controlling the AC Stark shift have led to a larger research effort concentrated on timescales above 10,000~s.

\section{Conclusions and Future Outlook} \label{Sec:Conclusion}

We have demonstrated our system is capable of averaging down less than $4.6\times 10^{-15}$ at 16,000~s.  While we have only limited experimental data for timescales beyond this, we have seen that the clock instability increases on longer timescales. It appears that this performance degradation is  related to the AC Stark shift, which would indicate that tighter control and better measurement of the laser power is required to achieve fractional frequency instabilities of $1\times10^{-15}$ at one day.   
The initial optical design of the vapor cell assembly implemented a photodiode for use in laser power stabilization that was thermally anchored to temperature stage 2 for reduction of temperature influenced drifts.  In some respects, the PMT used for fluorescence detection offers an improved measure of the average laser power across the atomic cloud because it relies on an atomic based signal rather than a beam sampling optic, for which the reflectivity is subject to polarization and temperature variations.  
 
Nevertheless, thermal considerations are still required, as the PMT's conversion efficiency is also temperature dependent \cite{Young1963}, thought to be believed to be somewhere between 0.5~\% to 1~\% per K.  Within the current design of the vapor cell assembly, an attempt was made to thermally connect the PMT to temperature stage~2, and fiberglass insulation was used to reduce thermal parasitics.  However, temperature gradients are difficult to minimize over the 10~cm length of the PMT. Additionally, the PMT's high voltage power supply and transimpedance amplifier need to provide constant voltage and gain, respectively, in order to use the PMT as a long-term laser power sensor. 
Alternatively, if tighter control of laser power proves too challenging, there are several design modifications that allow the two-photon clock to operate with a smaller Stark shift. First, the spectral linewidth could be decreased by removing the helium contamination in the vapor cell; in our current cells, which are expected to be saturated with 4~mTorr of helium, the linewidth is approximately a factor of 2 larger than the natural linewidth. Removing the helium (through the use of a vacuum chamber or a helium-resistant cell \cite{Dellis2016}) would allow for a reduced laser power by a factor of 2 without affecting the short-term stability. Additional modifications include  increasing the efficiency of the fluorescence detection; increasing the effective length of the vapor cell by routing the beam through the vapor on multiple non-overlapping passes; increasing the vapor density; and decreasing the desired short-term stability metric. Another final option is to increase the laser beam radius and laser power together; increasing both the power $P$ and the intensity radius $w_0$ by a factor of $F$ results in the same photon detection rate (accounting for both the decreased optical intensity and the increased number of atoms in the beam path) but a smaller overall AC Stark shift by the same factor $F$. 

Beyond improvements to the stability, future work with this system necessarily involves miniaturizing and hardening the laser and frequency comb for future field deployment. Particular emphasis will need to be paid to automation of the locking electronics and simplification of the control loop architecture, which currently includes 10 independent temperature control segments. Reducing the power consumption of the frequency comb, which currently uses four pump lasers, is another important area of research. Finally, any attempt to prepare this type of advanced optical clock for space will have to overcome the challenge of radiation-induced darkening of optical components. Despite these challenges, we remain optimistic that a vapor cell optical clock, such as the one described here, could soon be deployed to provide precise timekeeping for a host of applications.

\section*{acknowledgments} 
We thank Jordan Armstrong and the Space Dynamics Laboratory machine shop for assistance in constructing the experimental apparatus.  We thank Steve Lipson for careful reading of the manuscript.
\section{References}
%

\end{document}